\PassOptionsToPackage{table}{xcolor}
\documentclass[fleqn]{llncs}

\usepackage[firstpage]{draftwatermark}\SetWatermarkText{\hspace*{5.2in}\raisebox{5in}{\includegraphics[scale=0.1]{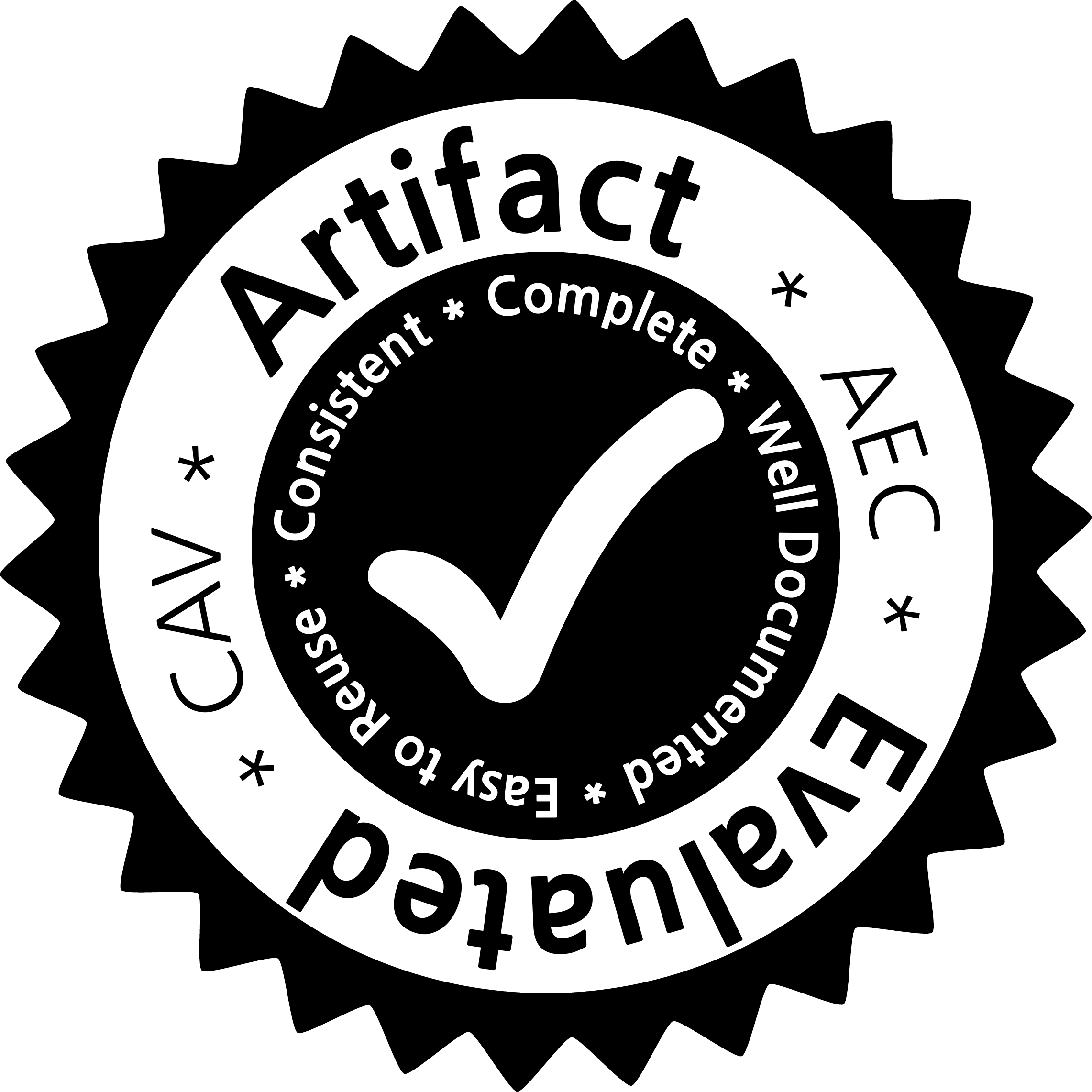}}}\SetWatermarkAngle{0}

\usepackage{verbatim}

\usepackage{amsmath}
\usepackage{amssymb}
\usepackage{stmaryrd} 
\usepackage{wasysym}
\usepackage{mathtools} 
\usepackage{bm}

\usepackage{multirow}
\usepackage{arydshln}

\usepackage{cite}
\usepackage{hyperref}

\usepackage{tikz}
\usetikzlibrary{arrows,automata,positioning}
\usetikzlibrary{decorations.pathmorphing}
\usepgflibrary{shapes.geometric}
\usepackage{pgfplots}

\usepackage{subfig}

\makeatletter
\newbox\sf@box
 {\def\sf@one{#1}%
  \def\sf@two{#2}%
  \setbox\sf@box\hbox
    \bgroup}%
 {  \egroup
  \ifx\@empty\sf@two\@empty\relax
    \def\sf@two{\@empty}
  \fi
  \ifx\@empty\sf@one\@empty\relax
    \subfloat[\sf@two]{\box\sf@box}%
  \else
    \subfloat[\sf@one][\sf@two]{\box\sf@box}%
  \fi}
\makeatother

\newcommand{\tool}[1]{\textsf{#1}}

\newcommand{\bosy}{\tool{BoSy}}

\newcommand{\ltlthreeba}{\tool{ltl3ba}}
\newcommand{\spot}{\tool{spot}}
\newcommand{\spin}{\tool{SPIN}}

\newcommand{\picosat}{\tool{PicoSAT}}
\newcommand{\cmsat}{\tool{CryptoMiniSat}}

\newcommand{\rareqs}{\tool{RAReQS}}
\newcommand{\depqbf}{\tool{DepQBF}}
\newcommand{\caqe}{\tool{CAQE}}
\newcommand{\quabs}{\tool{QuAbS}}
\newcommand{\bloqqer}{\tool{Bloqqer}}
\newcommand{\cadet}{\tool{CADET}}

\newcommand{\idq}{\tool{iDQ}}

\newcommand{\zthree}{\tool{Z3}}
\newcommand{\cvcfour}{\tool{CVC4}}

\newcommand{\nusmv}{\tool{NuSMV}}

\newcommand{\runsolver}{\tool{runsolver}}

\title{BoSy: An Experimentation Framework for Bounded Synthesis\thanks{Supported by the European Research Council (ERC) Grant OSARES (No.\ 683300).}}

\author{Peter Faymonville \and Bernd Finkbeiner \and Leander Tentrup}

\institute{Saarland University, Saarbr\"ucken, Germany\\\email{lastname@react.uni-saarland.de}}

\begin{document}

\maketitle

\begin{abstract}
We present $\bosy$, a reactive synthesis tool based on the bounded synthesis approach.  Bounded synthesis ensures the minimality of the synthesized implementation by incrementally increasing a bound on the size of the solutions it considers.
For each bound, the existence of a solution is encoded as a logical constraint solving problem that is solved by an appropriate solver.
$\bosy$ constructs bounded synthesis encodings into SAT, QBF, DQBF, EPR, and SMT, and interfaces to solvers of the corresponding type.  When supported by the solver, $\bosy$ extracts solutions as circuits, which can, if desired, be verified with standard hardware model checkers.
$\bosy$ won the LTL synthesis track at SYNTCOMP 2016. In addition to its use as a synthesis tool, $\bosy$ can also be used as an experimentation and performance
evaluation framework for various types of satisfiability solvers.

\end{abstract}

\section{Introduction}

The reactive synthesis problem is to check whether a given $\omega$-regular specification, usually presented as an LTL formula, has an implementation, and, if the answer is yes, to construct such an implementation. As a theoretical problem, reactive synthesis dates back all the way to Alonzo Church's solvability question~\cite{Church1963-CHUAOR} in the 1950s; as a practical engineering challenge, the problem is fairly new. Tools for reactive synthesis started to come out around 2007~\cite{conf/fmcad/JobstmannB06,conf/cav/FiliotJR09,conf/cav/BohyBFJR12,conf/tacas/Ehlers11}. The first SYNTCOMP tool competition took place at CAV 2014 and was originally restricted to safety specifications, and only later, starting with CAV 2016, extended with an LTL synthesis track~\cite{journals/corr/JacobsBBK0KKLNP16}. 

In this paper, we present $\bosy$, the winner of the 2016 LTL synthesis track. $\bosy$ is based on the bounded synthesis approach~\cite{journals/sttt/FinkbeinerS13}. Bounded synthesis ensures the minimality of the synthesized implementation by incrementally increasing a bound on the size of the solutions it considers.
For each bound, the existence of a solution is encoded as a logical constraint solving problem that is solved by an appropriate solver. If the solver returns ``unsat'', the bound is increased and a new constraint system is constructed; if the solver returns a satisfying assignment, an implementation is constructed.

From an engineering perspective, an interesting feature of the bounded synthesis approach is that it is highly modular. The construction of the constraint system
involves a translation of the specification into an $\omega$-automaton. Because the same type of translation is used in model checking, a lot of research has
gone into optimizing this construction; well-known tools include $\ltlthreeba$~\cite{conf/tacas/BabiakKRS12} and $\spot$~\cite{conf/atva/Duret-LutzLFMRX16}.
On the solver side, the synthesis problem can be encoded in a range of logics, including boolean formulas (SAT), quantified boolean formulas (QBF), dependency quantified boolean formulas (DQBF), the effective propositional fragment of first-order logic (EPR), and logical formulas with background theories (SMT) (cf.~\cite{encodings}). For each of these encodings, there are again multiple competing solvers.

$\bosy$ leverages the best tools for the LTL-to-automaton translation and the best tools for solving the resulting constraint systems. In addition to its main purpose, which is the highly effective synthesis of reactive implementations from LTL specifications, $\bosy$ is therefore also an experimentation framework, which can be used to compare individual tools for each problem, and even to compare tools across different logical encodings. For example, the QBF encoding is more compact than the SAT encoding, because it treats the inputs of the synthesized system symbolically. $\bosy$ can be used to validate, experimentally, whether QBF solvers translate this compactness into better performance (spoiler alert: in our experiments, they do). Likewise, the DQBF/EPR encoding is more compact than the QBF encoding, because this encoding treats the states of the synthesized system symbolically. In our experiments, the QBF solvers nevertheless outperform the currently available DQBF solvers.

In the remainder of this paper, we present the tool architecture, including the interfaces to other tools, and report on experimental results\footnote{$\bosy$ is available online at \url{https://react.uni-saarland.de/tools/bosy/}.}.

\section{Tool Architecture}

\begin{figure}[htb!]
  \centering
  \begin{tikzpicture}[>=stealth',shorten >=1pt,auto,thick,scale=0.75,transform shape]
    \tikzstyle{state}=[rounded corners,rectangle split, rectangle split parts=2, inner sep=5pt, draw]
    \tikzstyle{comp}=[rounded corners, inner sep=8pt, fill=white, draw]
    \tikzstyle{art}=[inner sep=5pt, fill=white, draw, minimum height=18pt]
    
	 \coordinate (start) at (0,0);
  
   \draw[thick, fill=gray!20, rounded corners] (start) -- ++(12, 0) -- ++(0,-10) 
   -- ++ (-9,0) -- ++ (0,-4)  -- ++ (9,0) -- ++ (0,-3)
   -- ++ (-12,0) -- ++ (0,10) -- ++ (9,0) -- ++ (0,4) -- ++ (-9,0)
   -- cycle;
   
   \node (input) at (4.5,.8) {LTL, Signature, Mealy/Moore};
   \node[comp] (pre) at (4.5,-1.5) {Preprocessing};
   
   \node[art] (ltl) at (4.5,-3) {LTL};
   
   \node[comp,align=center] (ltl2aut) at (4.5,-5) {LTL to Automata\\ Translation\\ \scriptsize ltl3ba, spot};

   \node[art] (aut) at (4.5,-7) {Automata};
   
   \node (impl) at (7.5,-17.8) {Implementation ({\texttt{AIGER, SMV, DOT}}) };
    
   \draw (input) edge[->] (pre);
   \draw (pre) edge[->] (ltl);
   \draw (ltl) edge[->] (ltl2aut);
   \draw (ltl2aut) edge[->] (aut);

   \draw[dashed] (0,-3.8) rectangle (8.2,-6.2); 
   
   \node[comp] (enc) at (6,-8.5) {Encoding};
   
   \node[art] (smt) at (4,-10) {SMT};
   \node[art] (sat) at (6,-10) {SAT};
   \node[art] (qbf) at (8,-10) {QBF};
   \node[art] (dqbf) at (10.5,-10) {DQBF / EPR};

   \draw (aut) edge[->] (enc);
   \draw (enc) edge[->] (smt);
   \draw (enc) edge[->] (sat);
   \draw (enc) edge[->] (qbf);
   \draw (enc) edge[->] (dqbf);

   \node[comp, align=center,inner sep=4pt] (smtsolver) at (4,-12) {SMT\\Solver\\ \scriptsize  z3, cvc4};
   \node[comp, align=center,inner sep=4pt] (satsolver) at (6,-12) {SAT\\Solver\\ \scriptsize picosat,\\ \scriptsize cmsat};
   \node[comp, align=center,inner sep=4pt] (qbfsolver) at (8,-12) {QBF Solver\\ \scriptsize rareqs, CAQE, \\ \scriptsize  depqbf, \\ \scriptsize QuAbs, CADET};
   \node[comp, align=center,inner sep=4pt] (dqbfsolver) at (10.7,-12) {DQBF / EPR\\Solver\\ \scriptsize idq / \\ \scriptsize eprover, vampire };
   
   \draw (smt) edge[->] (smtsolver);
   \draw (sat) edge[->] (satsolver);
   \draw (qbf) edge[->] (qbfsolver);
   \draw (dqbf) edge[->] (dqbfsolver);
  
   \node[art] (assignment) at (6,-14) {Assignment};
   \node[art] (certificate) at (8.5,-14) {Certificate};
   
   \draw (satsolver) edge[->] (assignment);
   \draw (qbfsolver) edge[->] (certificate);
   
   \draw[dashed] (3.2,-10.8) rectangle (12,-13.3); 
   
   \draw[->,dotted] (3.2,-12) .. controls (0,-12) and (0,-8) .. (enc.west);
   \node[align=center] at (1.5,-12.3) {unsatisfiable \\ increase bound};
   
   \node[comp] (post) at (7.5,-15.5) {Postprocessing};
   \draw (post) edge[->] (impl);
   \draw (assignment) edge[->] (post);
   \draw (certificate) edge[->] (post);

  \end{tikzpicture}
  \caption{Tool Architecture of $\bosy$}
  \label{fig:architecture}
  \vspace{-12pt}
\end{figure}
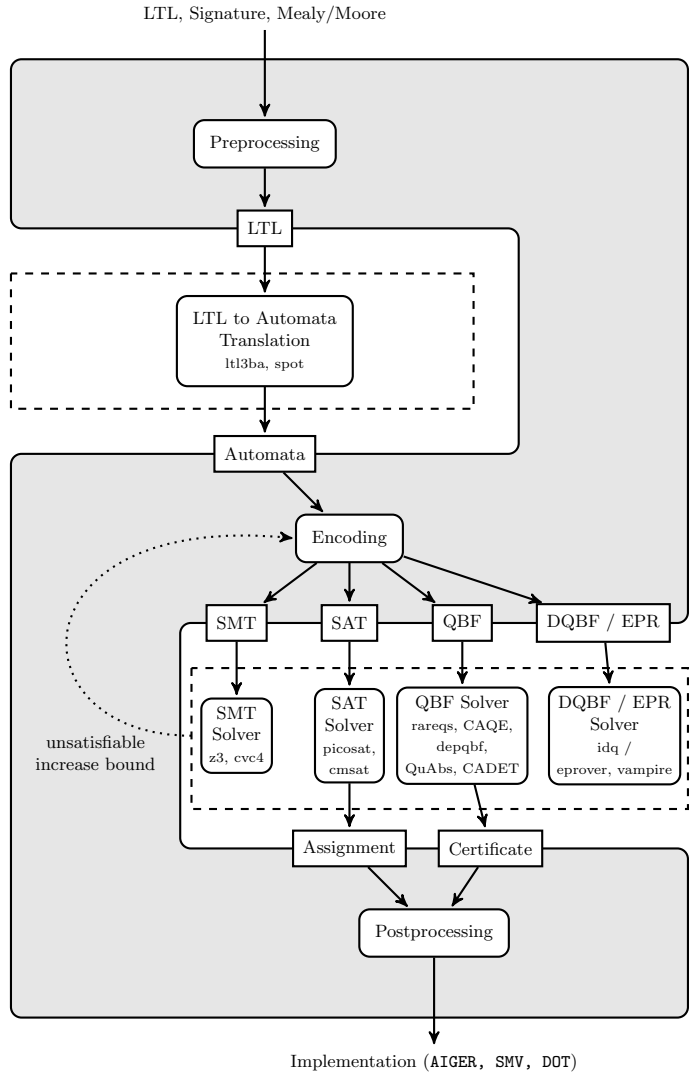

An overview of the architecture of $\bosy$ is given in Figure~\ref{fig:architecture}. For a given bounded synthesis instance, $\bosy$ accepts a JSON-based input format that contains a specification $\varphi$ given in LTL, the signature of the implementation given as a partition of the set of atomic propositions into inputs and outputs, and the target semantics as a Mealy or a Moore implementation. 
In the preprocessing component, the tool starts to search for a system \emph{strategy} with $\varphi$ and an environment \emph{counter-strategy} with $\neg \varphi$ in parallel.\footnote{The LTL reactive synthesis problem is not dual with respect to dualizing the LTL formula only, but the target semantics has to be adapted as well.
If one searches for a transition-labeled (Mealy) implementation, a counterexample is state-labeled (Moore) and vice versa.} 

After parsing, the LTL formula is translated into an equivalent universal co-B\"uchi automaton using an external automata translation tool. Currently, we support $\ltlthreeba$~\cite{conf/tacas/BabiakKRS12} and $\spot$~\cite{conf/atva/Duret-LutzLFMRX16} for this conversion, but further translation tools can be integrated easily. Tools that output $\spin$ never-claims or the HOA format are supported.

To the resulting automaton, we apply some basic optimization steps like replacing rejecting terminal states with safety conditions and an analysis of strongly connected components to reduce the size of the constraint system~\cite{journals/sttt/FinkbeinerS13}.

The encoding component is responsible for creating the constraint system based on the selected encoding, the specification automaton, and the current bound on the number of states of the system. The component constructs a constraint system using a logic representation which supports propositional logic, different kinds of quantification, and comparison operators (between natural numbers).
Our implementation contains the following encoding options. These encodings differ in their ability to support the symbolic encoding of the existence of functions.
We refer the reader to~\cite{encodings} for details.
\begin{itemize}
  \item A \emph{propositional} backend for SAT, where all functions are unrolled to conjunctions over their domain.
  \item An \emph{input-symbolic} encoding employing QBF solvers, where functions with one application context are symbolically represented.
  \item Two encodings (\emph{state-symbolic}, \emph{symbolic}) using DQBF/EPR solvers, where functions, which are used in multiple contexts, are encoded symbolically.
  \item An \emph{SMT} encoding resembling the original bounded synthesis encoding~\cite{journals/sttt/FinkbeinerS13}.
\end{itemize}

This constraint system is then translated to a format that the selected solver understands, and the solver is called as an external tool. 
We support SAT solvers that accept the DIMACS input format and that can output satisfying assignments, currently $\picosat$~\cite{journals/jsat/Biere08} and $\cmsat$~\cite{conf/sat/SoosNC09}.
We have three categories of QBF solving tools: QDIMACS/QCIR solver that can output top-level assignments ($\rareqs$~\cite{journals/ai/JanotaKMC16}, $\caqe$~\cite{conf/fmcad/RabeT15}, and $\depqbf$~\cite{journals/jsat/LonsingB10}), QDIMACS preprocessors ($\bloqqer$~\cite{conf/cade/BiereLS11}), and certifying QDIMACS/QCIR solver that can provide boolean functions witnessing satisfiable queries ($\quabs$~\cite{journals/corr/Tentrup16}, $\cadet$~\cite{conf/sat/RabeS16}, and $\caqe$~\cite{conf/fmcad/RabeT15}).
For the remaining formats, i.e., DQDIMACS ($\idq$~\cite{conf/sat/FrohlichKBV14}), TPTP3, 
and SMT ($\zthree$~\cite{conf/tacas/MouraB08}, $\cvcfour$~\cite{conf/cav/BarrettCDHJKRT11}), we only require format conformance as witness extraction is not supported, yet.

After the selected solver with corresponding encoding has finished processing the query and reports \emph{unsatisfiable}, the \emph{search strategy} determines how the bound for the next constraint encoding is increased. Currently, we have implemented a linear and an exponential search strategy.
In case the solver reports \emph{satisfiable}, the implementation will be extracted in the postprocessing component. The extraction depends on the encoding and solver support, we currently support it for SAT and QBF.

In case of the encoding to SAT, the solver delivers an assignment, which is then translated to our representation of the synthesized implementation.  The transition function and the functions computing the outputs are represented as circuits.

In case of the QBF-encoding, we take a two-step approach for synthesis.
The QBF query has the quantifier prefix $\exists \forall \exists$~\cite{encodings}.
In synthesis mode, the query is solved by a combination of QBF preprocessor and QBF solver.
From a satisfiable query, the assignment of the top-level existential quantification is extracted~\cite{conf/date/SeidlK14} and then used to reduce the original query by eliminating the top level existential quantifier.
The resulting query, now with a $\forall \exists$ prefix, is then solved using a certifying QBF solver that returns a certificate, that is a circuit representing the witnessing boolean functions.
This certificate is then translated into the same functional representation of the synthesized implementation as in the SAT case.

This common representation of the implementations allows the translation into different output formats.
From our representation, it is possible to translate the implementation into an AIGER circuit as required by the SYNTCOMP rules, to a SMV model for model checking, or to a graphical representation using the DOT format.

Further encodings can be integrated as extra components in the tool. Such an encoder component has access to the automaton, the semantics, and the input and output signals. The encoder must provide a method \texttt{solve} that takes as its parameter a bound and returns whether there is an implementation with the given bound that realizes the specification.
The method is implemented by building the constraint system and solving it using a theory solver.
One can either use our logic representation or build the textual solver representation directly.
If the component supports synthesis, it implements a second method \texttt{extractSolution} that is called if \texttt{solve} returns true.
It returns a representation of the realizing implementation as described above.
In order to integrate new solver formats, one has to provide a translator from our logic representation to this format.

\section{Experimentation}

The reactive synthesis competition has a library of LTL benchmarks that can be transformed into the $\bosy$ file format using the organizers' conversion tool~\cite{journals/corr/Jacobs016}.
The tool $\runsolver$~\cite{journals/jsat/Roussel11} is used in our experiments to get predictable timing results and to set appropriate time and memory limits.
Figure~2 compares the performance of the different encodings for determining realizability on the SYNTCOMP benchmark set.
Notably, the encoding employing QBF solving performed better than the SAT-based one and both solve more instances than the original SMT encoding.
The two encodings using DQBF are not yet competitive due to limited availability of DQBF solvers.

\begin{figure}[!tb]
  \centering
  \begin{tikzpicture}
    \begin{semilogyaxis}[xlabel=\# instances,ylabel=time (sec.),tiny,width=.75\columnwidth,height=7cm,mark size=1.7pt,ymin=0,ymax=3600,xmin=0,xmax=150,legend entries={symbolic (DQBF),state-symbolic (DQBF),SMT,propositional (SAT),input-symbolic (QBF)},
        legend style={
          at={(-0.1,0.95)},
          anchor=north east}]
        ]]
      \addplot+[brown,solid,mark=triangle*] table {plots/bosy-paper_cactus_bosy-realizability-fully-symbolic-exponential-g0.dat};
      \addplot+[orange,solid,mark=asterisk] table {plots/bosy-paper_cactus_bosy-realizability-state-symbolic-exponential-g0.dat};
      \addplot+[green,solid,mark=diamond*] table {plots/bosy-paper_cactus_bosy-realizability-smt-exponential-g0.dat};
      \addplot+[red,solid,mark=square*] table {plots/bosy-paper_cactus_bosy-realizability-non-symbolic-exponential-g0.dat};
      \addplot+[blue,solid,mark=*] table {plots/bosy-paper_cactus_bosy-realizability-input-symbolic-exponential-new-g0.dat};

    \end{semilogyaxis}
  \end{tikzpicture}
  \caption{Number of solved instances within 1 hour among the 195 instances from SYNTCOMP~2016. The time axis has logarithmic scale. The experiment was run on a machine with a $3.6\,\text{GHz}$ quad-core Intel Xeon processor.}
  \label{fig:cactus_solved_instance}
    \vspace{-12pt}
\end{figure}
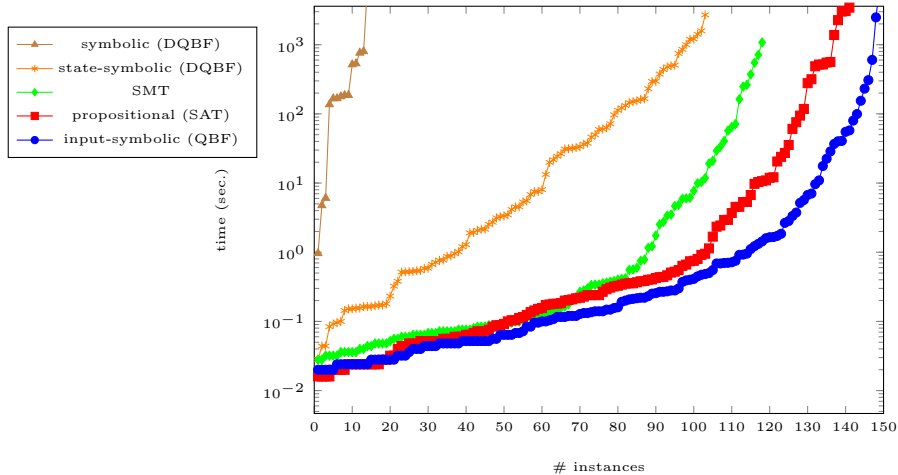


Different configurations of $\bosy$ may not only result in varying running times, but may also effect the \emph{quality} of the synthesized implementation.
A measurement of the quality of the implementation that has also been used in the reactive synthesis competition is the size of the implementation, measured in the number of gates in the circuit representation.
The correctness of an implementation, i.e., whether the synthesized solution actually satisfies the original specification can be verified by model checking.
One can either encode the solution as a circuit and use an AIGER model checker, or one can use the SMV representation and model check it with $\nusmv$~\cite{conf/cav/CimattiCGGPRST02}.

\begin{figure}[t]
  \centering
  \begin{tikzpicture}[->,>=stealth',shorten >=1pt,auto,thick,scale=0.7,transform shape]
    \tikzstyle{state}=[rounded corners,rectangle split, rectangle split parts=2, inner sep=5pt, draw,align=center]
    
    \coordinate (realizability_start) at (1.4,4.4);
    \coordinate (realizability_end) at (5.8,-4);
    \draw[rounded corners=4pt,fill=black!5] (realizability_start) rectangle (realizability_end);
    \node[below right=0 and 0 of realizability_start] {{\small realizability}};
    
    \coordinate (synthesis_start) at (-1.4,4.4);
    \coordinate (synthesis_end) at (-7.5,-4);
    \draw[rounded corners=4pt,fill=black!5] (synthesis_start) rectangle (synthesis_end);
    \node[below left=0 and 0 of synthesis_start] {{\small synthesis}};
    
    \node[state] (instance) {\textbf{load\_balancer} \nodepart{two} 5 clients\\Mealy};
    
    \node[state,above=0.5 of instance] (ltl3ba) {\textbf{ltl3ba} \nodepart{two} 98 states};
    \node[state,below=0.5 of instance] (spot) {\textbf{spot} \nodepart{two} 27 states};
    
    \node[state,above right=-0.5 and 1 of spot] (spot_propositional_real) {\textbf{propositional} \nodepart{two} $\picosat$: 53\,sec. \\ $\cmsat$: 177\,sec.};
    \node[state,below=0.2 of spot_propositional_real] (spot_input_sym_real) {\textbf{input-symbolic} \nodepart{two} $\rareqs$: 7\,sec. \\ $\caqe$: 7\,sec.};
    
    \node[state,above left=-0.5 and 1 of spot] (spot_propositional_synt) {\textbf{propositional} \nodepart{two} $\picosat$: 59\,sec., 728 gates \\ $\cmsat$: 192\,sec, 1171 gates};
    \node[state,below=0.2 of spot_propositional_synt] (spot_input_sym_synt) {\textbf{input-symbolic} \nodepart{two} $\quabs$: 66\,sec., 174 gates \\ $\caqe$: 66\,sec., 412 gates};
    
    \node[state,above right=-0.5 and 1 of ltl3ba] (ltl3ba_propositional_real) {\textbf{propositional} \nodepart{two} $\picosat$: 548\,sec. \\ $\cmsat$: 1055\,sec.};
    \node[state,below=0.2 of ltl3ba_propositional_real] (ltl3ba_input_sym_real) {\textbf{input-symbolic} \nodepart{two} $\rareqs$: 13\,sec. \\ $\caqe$: 22\,sec.};
    
    \node[state,above left=-0.5 and 1 of ltl3ba] (ltl3ba_propositional_synt) {\textbf{propositional} \nodepart{two} $\picosat$: 655\,sec., 997 gates \\ $\cmsat$: 1061\,sec., 1224 gates};
    \node[state,below=0.2 of ltl3ba_propositional_synt] (ltl3ba_input_sym_synt) {\textbf{input-symbolic} \nodepart{two} $\quabs$: 611\,sec., 254 gates \\ $\caqe$: 663\,sec., 1921 gates};
    
    \draw (instance) edge (ltl3ba)
          (instance) edge (spot)
          
          (spot) edge (spot_propositional_real)
          (spot) edge (spot_propositional_synt)
          (spot) edge (spot_input_sym_real)
          (spot) edge (spot_input_sym_synt)
          
          (ltl3ba) edge (ltl3ba_propositional_real)
          (ltl3ba) edge (ltl3ba_propositional_synt)
          (ltl3ba) edge (ltl3ba_input_sym_real)
          (ltl3ba) edge (ltl3ba_input_sym_synt)
          ;
    
  \end{tikzpicture}
  \caption{The diagram shows the result on solving time and implementation quality for different configurations of $\bosy$ on a sample specification.}
  \label{fig:load-balancer}
    \vspace{-12pt}
\end{figure}
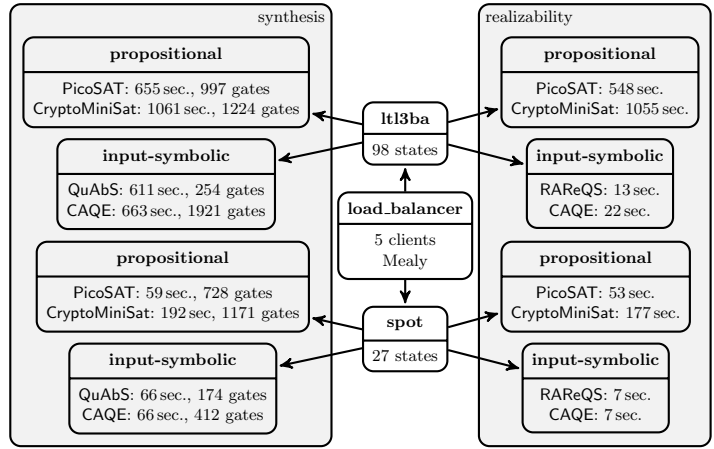

For a sample benchmark, a parametric load-balancer~\cite{conf/tacas/Ehlers11} instantiated with 5 clients, we provide experimental results for different configurations of $\bosy$ in Fig.~\ref{fig:load-balancer}.
The two automaton conversion tools produce significantly different automata, where the state space of the automaton produced by $\spot$ is only one third of the one produced by $\ltlthreeba$.
Consequently, the constraint system generated for the $\ltlthreeba$ version is larger and thereby the running time worse compared to the $\spot$ version.
This impact is stronger on the propositional encoding than on the input-symbolic one for realizability.
An observation that also translates to other benchmarks is that the size of the implementation is usually smaller using the input-symbolic encoding.
On the other hand, extracting solutions is cheaper in the propositional case as only assignments are extracted.

\bibliographystyle{splncs03}
\bibliography{main}

\end{document}